\documentclass[showpacs,twocolumn,prl]{revtex4}

\usepackage{amsfonts}
\usepackage{amssymb}
\usepackage{graphicx}
\usepackage{dcolumn}
\usepackage{bm}
\begin{document}

\title{Topological insulators in filled skutterudites}
\author{Binghai Yan$^{1}$, Lukas M\"{u}chler$^{1,2}$, Xiao-Liang
Qi$^{1}$, Shou-Cheng Zhang$^{1}$ and Claudia Felser$^{2}$ }

\affiliation{$^1$Department of Physics, McCullough Building,
Stanford University,Stanford, CA 94305-4045, USA \\
$^2$Institut f\"ur Anorganische Chemie und Analytische Chemie,
Johannes Gutenberg - Universtit\"{a}t,  55099 Mainz, Germany}

\begin{abstract}

We propose new topological insulators in cerium filled skutterudite
(FS) compounds based on \textit{ab initio} calculations. We find
that two compounds CeOs$_4$As$_{12}$ and CeOs$_4$Sb$_{12}$ are zero
gap materials with band inversion between Os-$d$ and Ce-$f$
orbitals, which are thus parent compounds of two and
three-dimensional topological insulators just like bulk HgTe. At low
temperature, both compounds become topological Kondo insulators,
which are Kondo insulators in the bulk, but have robust Dirac
surface states on the boundary. This new family of topological
insulators has two advantages compared to previous ones. First, they
can have good proximity effect with other superconducting FS
compounds to realize Majarona fermions. Second, the
antiferromagnetism of CeOs$_4$Sb$_{12}$ at low temperature provides
a way to realize the massive Dirac fermion with novel topological
phenomena.

\end{abstract}

\maketitle


The topological insulator (TI) is a new state of quantum matter
attracting extensive
attention\cite{qi2010,moore2010,hasan2010,qi2010RMP} for both
fundamental physics and technical applications recently. TIs have
insulating energy gap in the bulk and accommodate gapless edge or
surface states which are protected by time-reversal symmetry (TRS).
With TRS broken on the surface of a three-dimensional (3D)
insulator, the effective electromagnetic response is described by
the topological term $S_{\theta}=(\theta/2\pi)(\alpha/2\pi)\int
\mathrm{d}^3x\mathrm{d}t\mathbf{E}\cdot\mathbf{B}$, with $\alpha$
the fine-structure constant and the parameter $\theta=0$ or $\pi$
for trivial insulator and TI, respectively \cite{qi2008}. This
topological response supports many novel topological phenomena, such
as image magnetic monopole induced by a point charge\cite{qi2009},
topological Faraday and Kerr effects\cite{qi2008} and the
realization of axion field in condense matter physics\cite{li2010}.
Therefore magnetically doped TI for magnetic impurities and
ferromagnetism effects are of particular
interest. 
In addition, superconducting proximity effects on the surface states
of a 3D TI\cite{fu2008} 
have been proposed to realize the Majorana fermion which is a
promising candidate for quantum computation application. In this
letter we report new TI materials which could realize both the
antiferromagnetic (AFM) TI and the Majorana fermion. Furthermore
they are found to be topological Kondo insulators at low
temperature.

HgTe quantum wells were the first TI predicted
theoretically\cite{bernevig2006d} and subsequently observed
experimentally\cite{koenig2007}. The basic mechanism of band
inversion driven by spin-orbit coupling (SOC) was discovered in this
work and provides a template for most TIs discovered later.
Thereafter, many TIs have been proposed and experimentally measured,
e.g. Bi$_2$Se$_3$ and Bi$_2$Te$_3$
crystals\cite{zhang2009,xia2009,chen2009} and TlBiTe$_2$ \cite{yan2010,chen2010} 
crystals. Most of them are known as thermoelectric materials before.
At the same time theoretical calculations have predicted many other
TI materials which are waiting for experimental
verification\cite{hasan2010,qi2010RMP}. 
Here we find that CeOs$_4$As$_{12}$ and CeOs$_4$Sb$_{12}$ within the
filled skutterudite(FS) class are TI materials. The FS compounds
\cite{jeitschko1977} 
have a chemical formula $RT_4X_{12}$ ($R$= rare-earth, $T$= Fe, Ru
or Os, and $X$= P, As or Sb), in which heavy elements are expected
to induce strong SOC. Similar to Bi$_2$Se$_3$ materials, they are
also known for excellent thermoelectric properties
\cite{sales1996,Fleurial1996}
Moreover, they exhibit a rich variety of electronic and magnetic
ground states at low temperature\cite{sales2003,maple2008,sato2009},
including superconductivity, ferromagnetism, antiferromagnetism and
Kondo insulator behavior with hybridization gaps. Among these
compounds, most of Ce-based FSs are reported to be
insulators\cite{sales2003, maple2009}. All these reasons motivate us
to investigate the Ce-based FSs as TI candidates which can
accommodate magnetism and superconductivity proximity.

\begin{figure}
    \begin{center}
    \includegraphics[width=3.5 in]{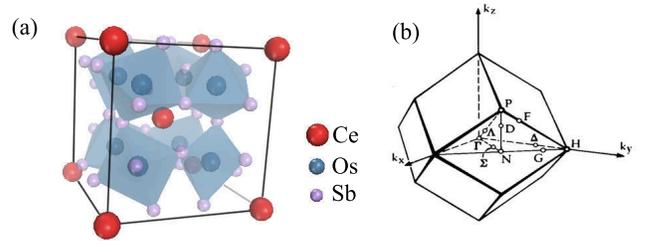}
    \end{center}
    \caption{ (Color online) (a) Crystal structures in $bcc$ lattice for filled skutterudites
$RT_4X_{12}$ ($R$= rare-earth, $T$= Fe, Ru or Os, and $X$= P, As or
Sb) in a $bcc$ lattice and (b) corresponding first Brillouin zone,
in which $\Gamma$, $N$ and $H$ are time-reversal-invariant points.}
    \label{fig:crystal}
\end{figure}


The FSs crystallize in the \textit{bcc} structure (Im3) shown in
Fig.1 and have inversion symmetry. The $R$ atoms, the inversion
centres, are located at the body centre and corners of the cubic
structure and surrounded by a cage of corner-sharing $T_4X_{12}$
octahedra. We calculated the electronic properties of Ce$T_4X_{12}$
by density-functional theory (DFT) within Perdew-burke-Ernzerbof
type generalized gradient approximation(GGA).
The \textit{Vienna ab initio simulation package}
(\textsc{vasp})\cite{kresse1993} with the projected augmented wave
method are employed. We adopted the experimental lattice constants
(see ref. \onlinecite{sales2003}) and fully optimized the atomic
positions in the unit cell. SOC is included in our calculations. In
addition, \textsc{wien2k} package\cite{blaha2001} was also used for
cross-checking.

\begin{figure*}
    \begin{center}
    \includegraphics[width=5.2 in]{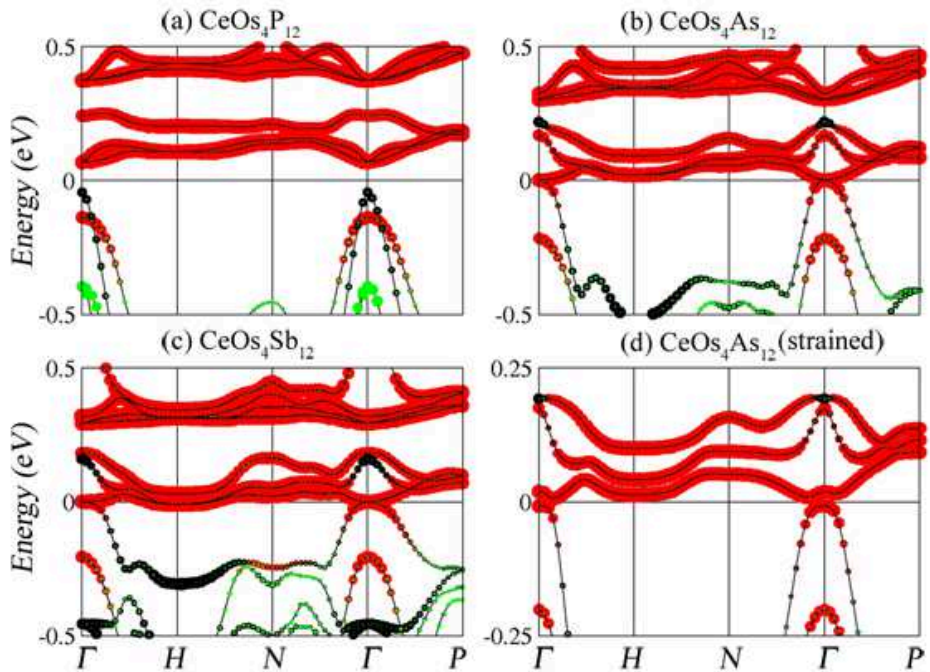}
    \end{center}
\caption{ (Color online) Band structures of (a) CeOs$_4$P$_{12}$,
(b)CeOs$_4$As$_{12}$, (c) CeOs$_4$Sb$_{12}$ and (d) strained
CeOs$_4$Sb$_{12}$. Red dots stand for the components of Ce-$f$
states, black dots for Os-$d$ states and green for $p$ states. The
size of dots represent the relative amplitude of corresponding
components. The Fermi energy is set to zero.}
    \label{fig:bands}
\end{figure*}

The calculated GGA band gap as well as the experimental values are
shown in Table I. Our results are well consistent with previous
calculations for CeFe$_4$P$_{12}$ and
CeFe$_4$Sb$_{12}$\cite{nordstrom1996} and
CeOs$_4$Sb$_{12}$\cite{harima2003}. Most materials have a small gap
while CeOs$_4$As$_{12}$ and CeOs$_4$Sb$_{12}$ have a zero gap.  We
show the band structure of CeOs$_4$X$_{12}$ in Fig.2. Take
CeOs$_4$P$_{12}$ as an example. The Ce-$4f$ states hybridize with
both Os-$5d$ and P-$3p$ states near the Fermi energy. A band gap
opens due to their hybridization. The hybridization can be
overestimated, since GGA tends to underestimate the inter-atomic
correlation of Ce-$4f$ electrons. Then the calculated band gap is
expected to be overestimated\cite{nordstrom1996}, which is different
from the usual gap underestimation of DFT calculation. This can
explain why our calculated band gap is larger than the experimental
one. For all these three FSs in Fig.2, the Ce-$f$ states exists near
the valence top and becomes dominant as the narrow spin-orbit
splitting bands at the conduction bottom.

\begin{table}
\caption{\label{table:bandgap} The calculated band gap and the
$\mathbb{Z}_2$ topological index. We listed the calculated band gap
$E_g$(calc.) together with the experimental value  $E_g$(exp.)(see
ref.\onlinecite{sato2009}) in unit of eV. The products of parities
for time-reversal invariant points in the $bcc$ Brillouin zone are
shown, including the $\Gamma$ point and one $H$ point and six $N$
points.}
\begin{ruledtabular}
\begin{tabular}{c|cccc}
FSs        & $E_g$(calc.) & $E_g$(exp.) & Parity &$\mathbb{Z}_2$ \\
\hline
CeFe$_4$P$_{12}$  & 0.38  & 0.13    &            &\\
CeFe$_4$As$_{12}$ & 0.16  & 0,0.01  &            & \\
CeFe$_4$Sb$_{12}$ & 0.08  & 0       &$\Gamma$(-) & \\
CeRu$_4$P$_{12}$  & 0.12  & 0.086   &$H$(-)      & (0;000)  \\
CeRu$_4$As$_{12}$ & 0.13  & 0,0.0043&$N$(-)      &  \\
CeRu$_4$Sb$_{12}$ & 0.12  & 0       &            & \\
CeOs$_4$P$_{12}$  & 0.12  & 0.034   &            & \\
\hline
                  &       &         & $\Gamma$(+)     &     \\
CeOs$_4$As$_{12}$ & 0.00  & 0.0047  & $H$(-)     &   (1;000) \\
CeOs$_4$Sb$_{12}$ & 0.00  & 0.0009  & $N$(-)     &    \\
\end{tabular}
\end{ruledtabular}
\end{table}

In order to determine their topological features, we use the parity
criteria proposed by Fu and Kane\cite{fu2007a} to calculate the
$\mathbb{Z}_2$ topological index.  The $\mathbb{Z}_2$ index is
determined by the parity of occupied bands on each time-reversal
invariant momenta. The $bcc$ Brillouin zone has eight
time-reversal-invariant points, including the $\Gamma$ point, six
$N$ points equivalent to ($\pi$,0,0) or ($\pi$,$\pi$,0) by point
group symmetry, and one $H$ point ($\pi$,$\pi$,$\pi$). We listed the
product of parities for all occupied states at these $k$-points and
corresponding $\mathbb{Z}_2$ index in Table I. Among them FSs
CeOs$_4$As$_{12}$ and CeOs$_4$Sb$_{12}$ are found to be topological
nontrivial with $\mathbb{Z}_2$(1;000). Compared to those topological
trivial FSs, e.g. CeOs$_4$P$_{12}$, a band inversion between Os-$d$
($\Gamma_5^+$) and Os-$f$ ($\Gamma_{67}^-$) bands is found in the
band structures, as shown in Fig. 2. The $d$($\Gamma_{5}^+$) state
enters the conduction bands. Both the valence band maximum and
conduction band minimum  at $\Gamma$ are $f$ ($\Gamma_{67}^-$)
states and touch with each other. The degeneracy induces a zero band
gap. This is similar with HgTe which has $s$ ($\Gamma_6$)-$p$
($\Gamma_8$) band inversion.In order to understand the band
structure better, we showed in Fig.3 the Fermi surface at $E_F$=0.0
and -0.1 eV cases for CeOs$_4$Sb$_{12}$. At the Fermi surface, there
is a small electron pocket around $H$ point from $f$ bands and a
tiny hole pocket at $\Gamma$. For $E_F$ slightly below zero, there
is a hole packet around $\Gamma$ with a little anisotropy, which is
due to the top valence band.

\begin{figure}
    \begin{center}
    \includegraphics[width=3.5 in]{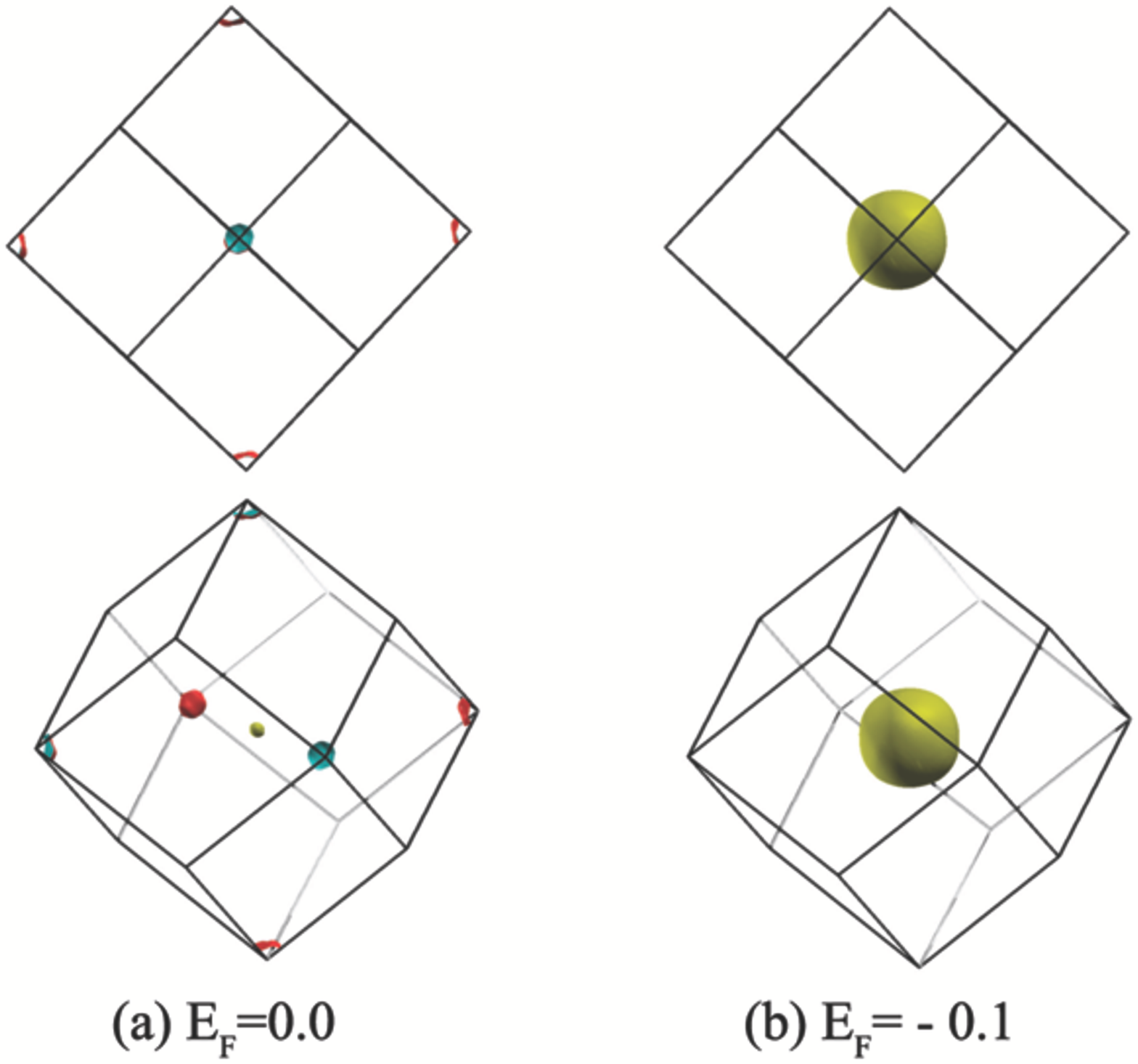}
    \end{center}
\caption{ (Color online) Fermi surface for CeOs$_4$Sb$_{12}$ at
$E_F=$ (a) 0.0 and (b) -0.1 eV. The top and side views are shown in
upper and lower panels, respectively.}
    \label{fig:bands}
\end{figure}

To drive CeOs$_4$As$_{12}$ and CeOs$_4$Sb$_{12}$ into real
topological insulators, the degeneracy at $\Gamma$ point needs to be
lifted. There are several possible ways to lift the degeneracy.
Firstly, it is possible to fabricate quantum wells using
CeOs$_4$As$_{12}$ or CeOs$_4$Sb$_{12}$ together with above
topological trivial FSs as barrier layers, like HgTe-CdTe quantum
wells\cite{bernevig2006d,koenig2007}. Then the 2D TI (quantum spin
hall effect) can be realized. Secondly, strain can lift up the
$\Gamma_{67}^-$ degeneracy and open a a gap at $\Gamma$. Then these
two materials will become 3D topological insulators, similar to
strained HgTe\cite{fu2007a,dai2008,bruene2011}. For example, 2\%
uniaxial strain along (001) direction can open an energy gap 
at $\Gamma$ for CeOs$_4$As$_12$ in our calculation, as shown in Fig.
2(d). Thirdly, a new feature in these two materials compared to
previously proposed TI materials is that they can be 3D topological
Kondo insulators\cite{dzero2010}. At low temperature ($T<135$ K for
CeOs$_4$As$_{12}$ and $T<50$ K for CeOs$_4$Sb$_{12}$), both
materials are reported to be Kondo
insulators\cite{bauer2001,baumbach2008}. Though the GGA band gap is
zero, at low temperature the residual carriers in the system can
form Kondo singlets with localized $f$ states, leading to a Kondo
insulator behavior in transport. Moreover, both materials have
$p$-type carrier\cite{sugawara2005,maple2011},  %
so that the Fermi
level does not cross the bands involved in the band inversion.
Consequently, when the carriers become insulating due to Kondo
effect, the topological surface state due to the band inversion are
expected to remain robust on the surface, so that the system becomes
a ``topological Kondo insulator" rather than ordinary Kondo
insulators. On the contrary, if the carrier is $n$-type, the bands
involved in band inversion will be buried below the Fermi level, so
that the topological nontrivial nature of the system is lost for
high enough carrier density. If the carrier density in this system
can be tuned, we expect to see a topological phase transition
between the topological Kondo insulator phase for $p$ doping and the
ordinary Kondo insulator phase for $n$ doping. It should be noticed
that the topological Kondo insulators studied here are different
from those proposed by Dzero {\it et al.}\cite{dzero2010}, since the
residual carrier density in the FS compounds is low, so that the
Kondo effect leading to a Kondo insulator state is expected to be
induced by magnetic impurities, rather than a Kondo lattice formed
by $f$-electrons. On a first view the Ce filled FSs have similar
electronic structure like the rare earth containing Heusler
compounds, which are also predicted to be HgTe-like TIs
recently\cite{chadov2010,lin2010a}. Among them the heavy-fermion  
behavior (e.g. YbPtBi) and superconductivity (e.g. LaBiPt) are
closely related to $f$ electrons. However, the topological feature
of the Heusler compounds are related to a $s-p$ band inversion,
different from these FSs.

Recent experiments reveal that CeOs$_4$Sb$_{12}$ has a AFM phase
below 1 K\cite{yang2007}. (It should be noted in above calculations
we constrained the magnetic moments as zero to keep the TRS, since
the total energy difference between with and without magnetic
moments is negligible small and beyond the accuracy of DF
calculations) The existence of AFM order provides a possible way to
realize the massive Dirac fermion state with many exotic topological
phenomena without requiring magnetic doping. For example, the spin
wave in the AFM phase is expected to have ``axionic" coupling to
photons, leading to the axion polariton effect and a new kind of
tunable optical modulator\cite{li2010}. Moreover, in proximity with
their superconducting FS neighbors, e.g. La$T_4X_{12}$, both
CeOs$_4$As$_{12}$ and CeOs$_4$Sb$_{12}$ can help to realize the
topological superconductivity as both 3D and 2D TIs.

In summary, we proposed two FS compounds CeOs$_4$As$_{12}$ and
CeOs$_4$Sb$_{12}$ as new topological insulators. Up to now the
experimentally measured TI materials include only $s$-$p$ band
inversion such as HgTe and $p$-$p$ band inversion such as
Bi$_2$Se$_3$ family. Different from them, these two TIs have $d$-$f$
band inversion. They have two advantages compared to previous ones.
Firstly, they can have good proximity with other superconducting FS
compounds to realize Majarona fermions. Secondly, the
antiferromagnetism of CeOs$_4$Sb$_{12}$ at low temperature provides
a way to realize the massive Dirac fermion with exotic topological
phenomena. Moreover, their rich physics offers a platform for
topological Kondo effects. At low temperature, both compounds become
topological Kondo insulators, i.e. they are Kondo insulators in the
bulk, but have robust Dirac surface states on the boundary. After
the completion of this work, we became aware of recent work on the
skutterudite CoSb$_3$\cite{pickett2011}, which is near the
topological critical point.

We acknowledge B. Maple and Y.L. Chen for the grateful discussion.
This work is supported by the Department of Energy, Office of Basic
Energy Sciences, Division of Materials Sciences and Engineering,
under contract DE-AC02-76SF00515 and by the Keck foundation. B. Y.
thanks the support by the Supercomputer Center of Northern Germany
(HLRN Grant No. hbp00002). L. M. gratefully acknowledges financial
support by the Graduate ``School of Excellence Material science in
Mainz'' (MAINZ).


\end{document}